# Time of Flight Modulation of Intensity by Zero Effort on Larmor

**N. Geerits[a,b], S. R. Parnell[a], M.A. Thijs[a], A.A. van Well[a], C. Franz[c,d], A.L. Washington[e], D. Raspino[e], R.M. Dalgliesh[e] and J. Plomp[a]**

[a]Faculty of Applied Sciences, Delft University of Technology, Mekelweg 15, Delft, 2629JB, The Netherlands

[b]Atominstitut, TU Wien, Stadionallee 2, Vienna, 1020, Austria

[c]Physik Department, Technische Universität München, Garching, D-85748, Germany

[d]Heinz Maier-Leibnitz Zentrum (MLZ), Technische Universität München, Garching, D-85748, Germany

[e]ISIS, Rutherford Appleton Laboratory, Chilton, Oxfordshire, OX11 0QX, UK

Correspondence email: j.plomp@tudelft.nl

**Abstract**   A time of flight Modulation of Intensity by Zero Effort spectrometer mode has been developed for the Larmor instrument at the ISIS pulsed neutron source. The instrument utilizes resonant neutron spin flippers which employ electromagnets with pole shoes, allowing the flippers to operate at frequencies of up to 3 MHz. Tests were conducted at modulation frequencies of 103 kHz, 413 kHz, 826 kHz and 1.03 MHz, resulting in a Fourier time range of ~0.1 ns to 30 ns using a wavelength band of 4 Å to 11 Å.

# 1. Introduction

Inelastic neutron scattering measures dynamics in materials by determining the energy transfer between the sample and the neutron beam. Resolution limitations occur in traditional techniques when the energy transfer is much smaller than the accuracy with which neutron energy can be determined (~µeV). To overcome this limitation, interferometric techniques based on the neutron spin-echo principle [1] can be used.

Neutron spin-echo instruments (NSE) [1, 2] employ polarised neutrons and two magnetic field regions. According to the semi-classical description, the first region will separate the two neutron spin states and the second region will recombine them again. Inelastic scattering from a sample occurs between the two field regions and induces a phase shift of the recombined states depending on the energy exchange of the scattering process. The distribution of the energy transfer will lead to a distribution of phases and will reduce the spin-echo amplitude for the neutron beam [3]. This principle leads to a high energy resolution in the order of neV [4]. Measurements of magnetic samples are possible [5,6], however it is challenging in cases where a sample depolarises the beam.

An alternative method to NSE, that is not affected by sample induced depolarisation, is the Modulation of IntEnsity by Zero Effort (MIEZE) technique (shown in figure 1) introduced by Gähler [8], which utilizes only one magnetic field region created by resonant spin flippers [9]. However, the energy resolution of MIEZE spectrometers (~300-600 neV or 1-2 ns [10-12] and in the case of longitudinal MIEZE [13,14] 30 neV or 20 ns) is significantly lower than that of NSE instruments (~0.2 neV). For several systems an energy resolution on the order of ~120 neV corresponding to a Fourier time of 5 ns is sufficient [15, 16] and in other cases longer Fourier time is required [17, 18]. To achieve the highest possible resolution in MIEZE with a suitable beam size, large and homogeneous magnetic fields are required in the resonant spin flippers, furthermore the RF field must oscillate at several MHz. These fields and frequencies are available on the Larmor instrument at the ISIS pulsed neutron source, a versatile Small Angle Neutron Scattering (SANS) instrument which is also capable of employing a variety of spin echo techniques such as Spin Echo SANS (SESANS) [19] and Spin Echo Modulated SANS (SEMSANS) [20]. This instrument utilizes spin flippers, which employ soft iron pole shoe magnets. The pole shoes enable the generation of homogeneous high magnetic fields at relatively low current. Furthermore, unlike most resonant spin flippers these flippers use longitudinal solenoids to generate an RF field parallel to the beam. Hence, they do not employ any material in the beam, which could induce scattering, reducing the intensity and q (scattering vector) resolution. In addition, due to the pulsed nature of the ISIS neutron source, the setup studied here is a time of flight MIEZE spectrometer [21], which yields additional advantages by utilising simultaneously the large wavelength band of neutrons.

## 2. Theory

The MIEZE arrangement is shown schematically in figure 1. The resonant spin flippers, like the magnetic field regions in NSE, introduce a longitudinal Stern-Gerlach effect. The first spin flipper introduces a kinetic energy shift between the two spin states, which causes the neutron spin states to move out of phase as they move towards the second spin flipper. The second flipper reverses this effect by overcompensating the kinetic energy shift, hence the phase difference will decrease towards the detector plane. Finally, once the neutrons reach the detector, the phase difference should be zero for every individual neutron. This is called the focusing plane. The energy difference between the two interfering spin states manifests itself as a beating of the polarisation in time [22, 23]. When an inelastic scattering sample is placed downstream of the second spin flipper the inelastic scattering process will influence both spin states in the same way. However, as the total kinetic energy has changed, the position of the focusing plane will move away from the detector plane. The distribution of scattering energies will yield a distribution in the focusing plane position and can be measured as a reduction of the depth of the modulation pattern. The amplitude of the MIEZE signal is therefore a measure of the energy transfer between the neutrons and the sample. The sample can be placed anywhere in the instrument after the first RF flipper, even downstream of the analyser, as its position will only affect the sensitivity, since the path length from sample to the detector dictates the magnitude of the shift of the focusing plane. Hence, with this technique all inelastic scattering, both incoherent and magnetic, is determined in one measurement. This also allows MIEZE to measure samples that depolarise the beam, [24] since sample induced spin rotations are not encoded when the sample is place downstream of the analyser.

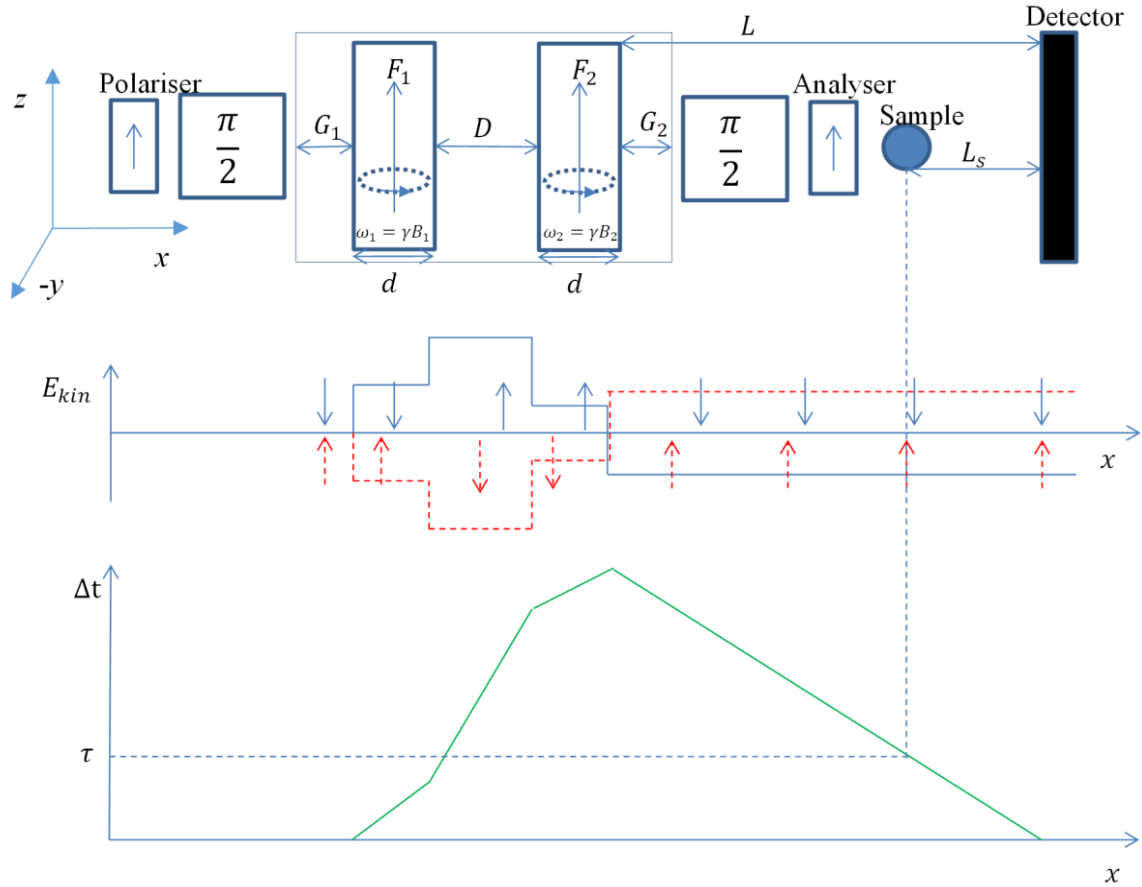

**Figure 1** *Schematic of the MIEZE spectrometer as implemented on the Larmor instrument (top). The flippers $F_1$ and $F_2$(thickness d) are positioned, a distance D apart, within a guide field. $\frac{\pi}{2}$ flippers are placed outside of the guide field to project the polarised neutrons on the y-axis. A sample can be placed downstream of the analyser. The semi-classical explanation (middle) shows the separation of kinetic energy introduced by the spin flippers. This splitting causes the two spin states, represented by the arrows, to move apart. The second resonant spin flipper $F_2$ overcompensates this energy splitting and ensures that the phase difference between the two spin states is zero at the detector position. Note that the arrows indicate the locations of the respective wavepacket and not the locations of equal phase. Inelastic scattering from a sample placed in the beam (a distance $L_s$ from the detector) will result in a reduction of the modulation depth seen at the detector. The corresponding Fourier time is shown at the bottom depending on the sample position. Adapted from reference 7.*

## 2.1. MIEZE Condition

In a conventional MIEZE setup the RF flippers are placed in a zero field chamber so no spurious fields will distort the encoding. As shown in figure 1, the time of flight MIEZE setup described in this work utilizes a guide field that needs to be taken into account. The MIEZE focus, derived in [22, 25], which indicates the detector position where the modulation of intensity is maximal, is shifted and the modified MIEZE condition is given by

$$L = \frac{2\omega_1 D + \omega_g(G_1 + G_2 - D)}{2\Delta\omega_m} + \frac{d}{2} \qquad (1)$$

Where $\Delta\omega_m = \omega_2 - \omega_1$, $\omega_g = \gamma B_g$ with $B_g$ the guide field strength. All other parameters are defined in figure 1. If this modified condition is fulfilled, the intensity at the detector will oscillate in time with a frequency equal to $2\Delta\omega_m$. This frequency is often referred to as the modulation frequency or MIEZE frequency. In the special case, where $D = G_1 + G_2$ and the guide field is homogenous, this modified expression is reduced to the classical MIEZE condition, given in [22]. This special case was used in our TOF Larmor setup as the position of the flippers can easily be tuned to this condition.

## 2.2. Intermediate Scattering Function

In the low energy transfer limit the ratio of the MIEZE amplitude with ($P_{sample}$) and without ($P_0$) sample yields the intermediate scattering function [1], which is the cosine Fourier transform of the scattering function:

$$\frac{P_{sample}}{P_0} = S(q,\tau) = \int_{-\infty}^{+\infty} S(q,\omega) \cos(\omega\tau)\, d\omega \qquad (2)$$

Where $\omega$ is the energy transfer and the Fourier time, is given by

$$\tau = \frac{\Delta\omega_m m^2 \lambda^3}{\pi h^2} L_s \qquad (3)$$

Where $m$ is the neutron mass, $L_s$ the sample to detector distance, $\lambda$ the neutron wavelength and $h$ is Plancks constant. Thus the Fourier time can be selected by tuning the instrument parameters such as the modulation frequency, the sample - detector distance or the neutron wavelength. In time-of-flight (TOF) MIEZE - a wavelength band is scanned in each TOF pulse; therefore a Fourier time range is scanned in a single pulse. Since the Fourier time is proportional to the reciprocal of the energy transfer, a large Fourier time corresponds to a good energy resolution.

## 2.3. TOF MIEZE Frequency Shift

TOF MIEZE has a rather unique property at its disposal, namely that the modulation frequency is shifted if the MIEZE condition (eq. 1) is not satisfied [25, 26]. In a monochromatic approach the violation of the MIEZE condition introduces a wavelength dependent phase shift to the modulation. In time-of-flight the neutron wavelength is time dependent, thus the wavelength dependent phase shift, manifests as a shift of the MIEZE frequency. Furthermore, due to the better wavelength resolution on time of flight setups compared to most monochromatic setups the modulation amplitude drops less quickly when the detector is moved out of focus.

In the case of a misaligned setup, where the detector is displaced by a distance $\Delta L$ from the focal point the expectation value of the neutron spin in the x-direction is given by the following expression

$$<\sigma_x> = \int_{-\infty}^{\infty} R(\lambda, \lambda_0, \Delta\lambda)\cos\left(2\Delta\omega_m\left[t - \frac{m\lambda\Delta L}{h}\right]\right)d\lambda = Re[F(R(\lambda, \lambda_0, \Delta\lambda))]\cos\left(2\Delta\omega_m\left[1 - \frac{\Delta L}{L}\right]t\right) \quad (4)$$

Where $R(\lambda, \lambda_0, \Delta\lambda)$ is the normalized wavelength resolution function, which is averaged over and $Re[F(R(\lambda, \lambda_0, \Delta\lambda))]$ indicates the real part of the Fourier transform. This is directly related to the envelope of the wavefunction and therefore the coherence length. The shifted frequency is thus given by

$$\omega' = 2\Delta\omega_m\left(1 - \frac{\Delta L}{L}\right) \quad (5)$$

This frequency shift simplifies the alignment procedure of the setup. Furthermore, it can yield additional information on non-symmetric inelastic scattering as the observed MIEZE frequency shift will have a different sign for energy gain or loss.

### 3. Experimental Setup

Figure 2 shows a graphical representation of the experimental setup where a double v-cavity polarizer and a single mirror analyser are used to polarise and analyse the polarisation of the beam. The incident beam has a circular shape at the sample position (diameter 20 mm). A gradient RF flipper (not depicted) is placed between the polarizer and the first π/2 flippers allowing sequential measurement of both spin states to determine the beam polarisation. The π/2 flippers (v-coils) [27, 28] are employed at the entrance and exit of the guide field, which create a superposition of the spin up and down state. A guide field (~0.6 mT) spanning from the edges of both v-coils surrounds the setup. The instrument is setup such that $D = G_1 + G_2$ (1.209m) and $G_1 = G_2 = \frac{D}{2}$. As a result once the MIEZE focus is found using the frequency shift technique, one can double, triple etc. the frequencies of the spin flippers, without displacing the focus.

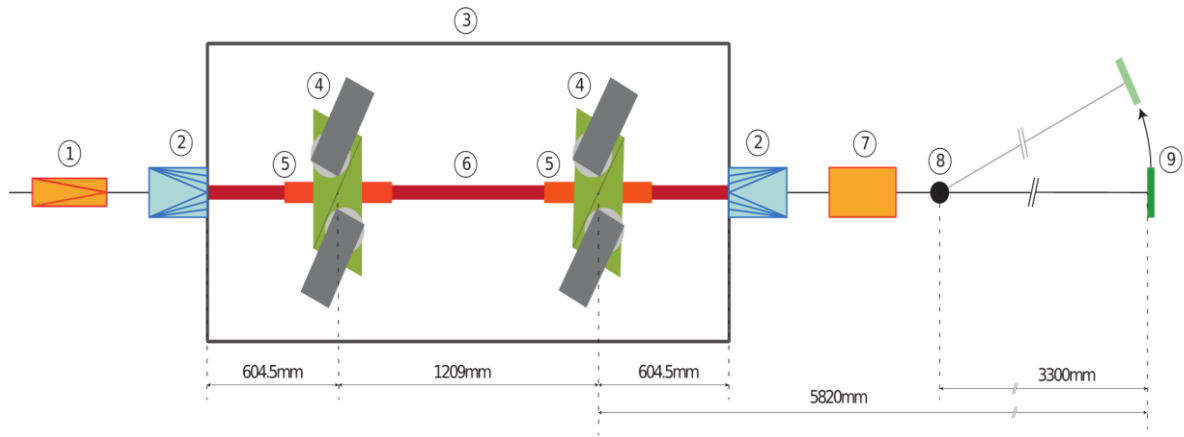

**Figure 2** *Schematic of the Larmor instrument in MIEZE mode, showing the most essential components: (1) double v-cavity polarizer, (2) v-coils, (3) guide field coil, (4) pole shoe magnets, (5) RF coils, (6) Flight tube, (7) Single mirror analyser, (8) Sample position (no samples were measured in these experiments) and (9) Detector.*

The spin flippers used on Larmor are shown in figure 3, these consist of a copper shielded longitudinal solenoid which generates the RF field and an electromagnet with soft iron pole shoes which allow static magnetic fields up to ~120 mT. The design of these RF flippers is very flexible as they can be operated in a gradient (or adiabatic) flipper mode [29, 27] and also a resonant flipper mode [9, 23]. The latter is used for the experiments described in this paper due to the lower power consumption and the large band of useable flipper frequencies using direct RF amplifier drive from 35 kHz to 3MHz. The low power consumption makes a resonant matching circuit obsolete, since the RF amplifiers are able to handle the reflected power. Thus, the amplifier outputs are connected directly to the RF solenoids. When used in time of flight mode the RF amplitude of a resonant flipper must be modulated using a 1/t function to match the exact π-flip for all wavelengths, where t is the time-of-flight from the source to the respective flipper. This ensures a high flipping efficiency [30] (>99%), over a wavelength range from 2.5 Å to 13.5 Å.

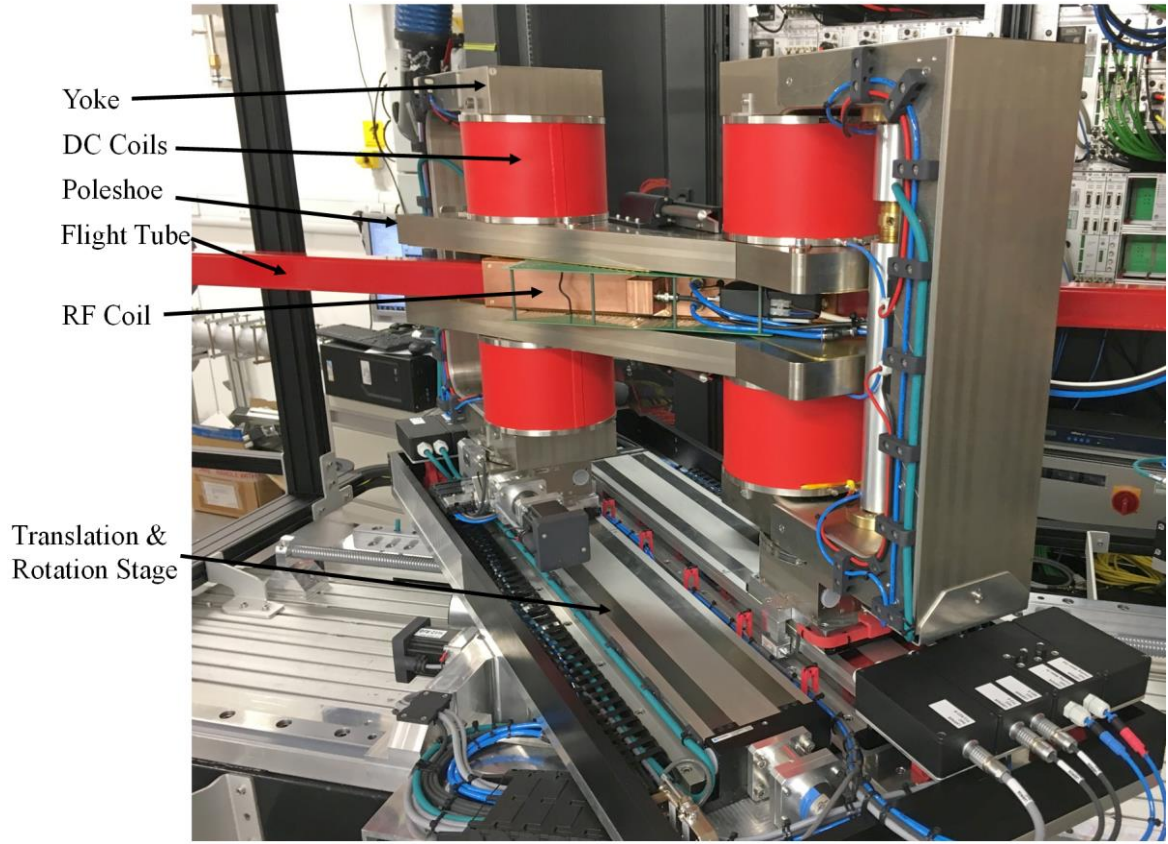

**Figure 3** *Picture of an RF spin flipper used by the Larmor instrument. This spin flipper has soft iron pole shoes to generate the required static magnetic field and a longitudinal RF coil to create the oscillating field. Furthermore the flipper has a gradient coil (not used in this experiment), which can be used to produce an adiabatic RF spin flip. The yoke pole shoe configuration can be translated and rotated on the table, thus allowing for both tilted and straight field regions. The latter is required for MIEZE, while the former is used in SESANS and SEMSANS. A vacuum exists within the flight tube to minimize air scattering.*

Each RF flipper is operated at four different frequencies (c.f. table I), giving rise to four different modulation frequencies. This results in a large Fourier time range at the sample position, which is 3.3 m from the detector, though no samples were used for these experiments. The wavelength range that could be used with reasonable statistics was 4 Å - 11 Å, with the peak of the spectrum being at roughly 4 Å.

**Table I** RF flipper frequencies and the resulting modulation frequency ($2\Delta f$).

| $f_1$ [MHz] | $f_2$ [MHz] | $2\Delta f$ [MHz] |
|---|---|---|
| 0.2484 | 0.3000 | 0.1032 |
| 0.9936 | 1.2000 | 0.4128 |
| 1.9872 | 2.4000 | 0.8256 |
| 2.4840 | 3.0000 | 1.0320 |

The detector is positioned 5.82 m from the second RF flipper. High frequency MIEZE setups require a detector with a fast response time and a thin detection path as this determines the time resolution. A Gas Electron Multiplier (GEM) type detector [31] was used, which employs a thin boron sheet where neutrons that are absorbed emit α particles [32, 33]. Such a detector has a sampling rate up to 100 MHz. Furthermore it has a 2D spatial sensitivity with a pixel size of 0.8 mm$^2$, allowing for a spatial analysis of the MIEZE signals, which is useful for determining the spatial field homogeneity of the spectrometer. Note that it is not possible to measure the instantaneous polarisation as it varies in time, rather the detector measures the average polarisation over the sampling time interval and the detector area. This averaged polarisation is equal to the visibility, which is defined as

$$V = \frac{I_+ - I_-}{I_+ + I_-} \tag{6}$$

with $I_+$ the intensity of the first spin state and $I_-$ the intensity of the second spin state [34]. Alternatively it can also be useful to look at the absolute value of the Fourier transform of the visibility, which is termed, spectral amplitude in this paper.

## 4. Results

Figure 4 shows the MIEZE signals for various modulation frequencies.

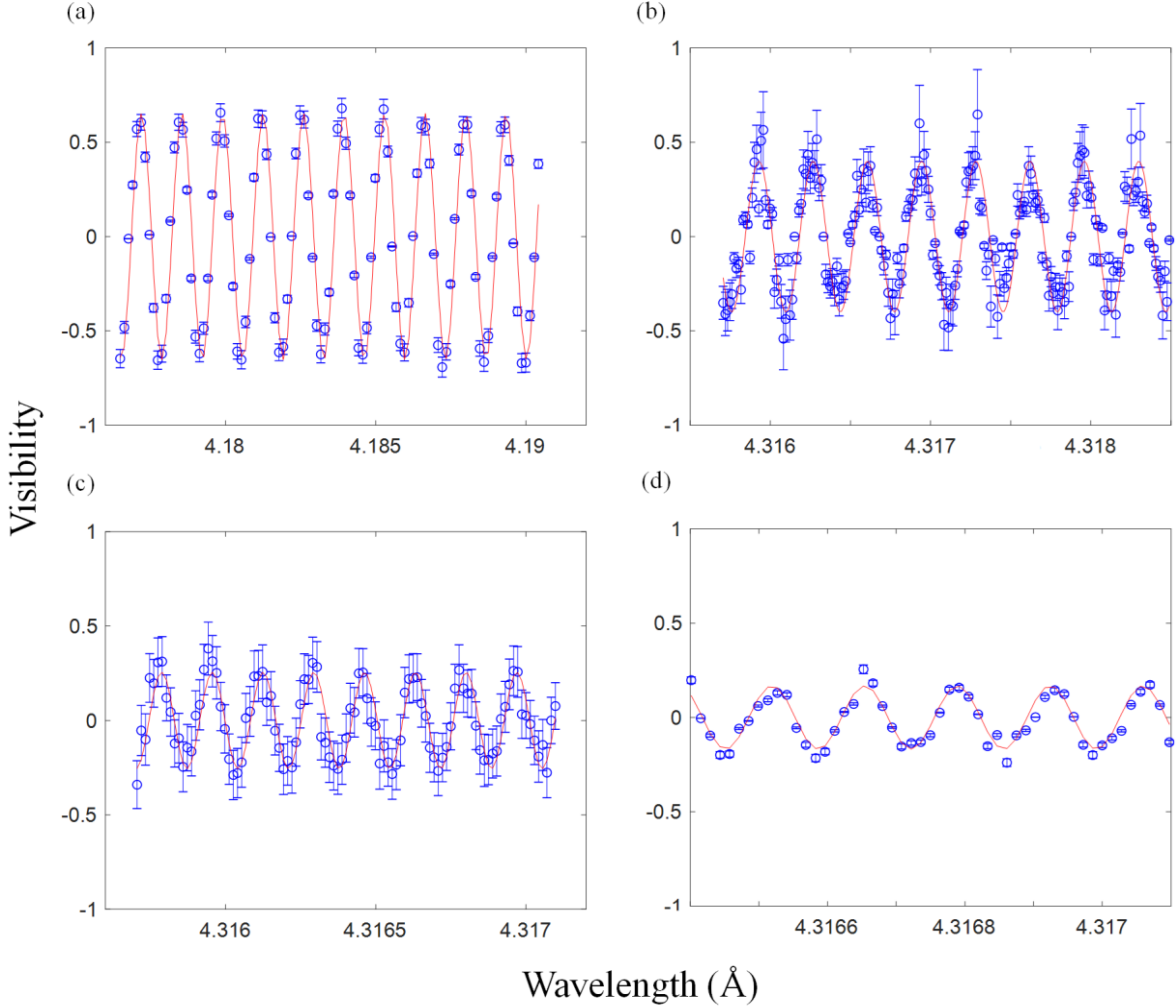

**Figure 4** *MIEZE signals at various frequencies with sinusoidal fits. The MIEZE frequencies are 103kHz (a), 413kHz (b), 826kHz (c) and 1.03MHz (d). The visibility shown here is the average visibility over a circular detector area with a diameter of 28 mm. Here, the modulation is shown in a limited wavelength range. The wavelength-dependent modulation can be found in figure 4.*

The 103 kHz signal was measured for a total of 60 minutes, the 413 kHz signal for 90 minutes, the 826 kHz signal for 105 minutes and the 1.03 MHz signal for 270 minutes. Measurement times are long due to the efficiency (~16% at 4 Å) of the GEM detector used which results from the thin detection volume. The respective sampling rates were 1 MHz for the 103 kHz signal and 10 MHz for the other signals. Further analysis of individual pixels demonstrated that there are no averaging effects, which negatively affect the visibility when averaging over the entire detector surface. Table II shows the measured modulation frequency against the expected modulation frequency.

**Table II** Expected modulation frequency, $f = \Delta\omega_m/\pi$ in MHz and the measured modulation frequency in MHz. The frequencies and their respective uncertainties are estimated from least squares fits to a sine function

| Expected (MHz) | 0.1032 | 0.4128 | 0.8256 | 1.0320 |
|---|---|---|---|---|
| Measured (MHz) | 0.1031±0.0014 | 0.4119±0.0005 | 0.8229±0.0067 | 1.0193±0.0337 |

Using a least squares regression the amplitude of the MIEZE signal is determined at various wavelengths (4 Å-11 Å for 1.03 MHz and 4 Å-8 Å for the others). This allows an extraction of the visibility as a function of the Fourier time at our Larmor sample position shown in Figure 5 for all four modulation frequencies.

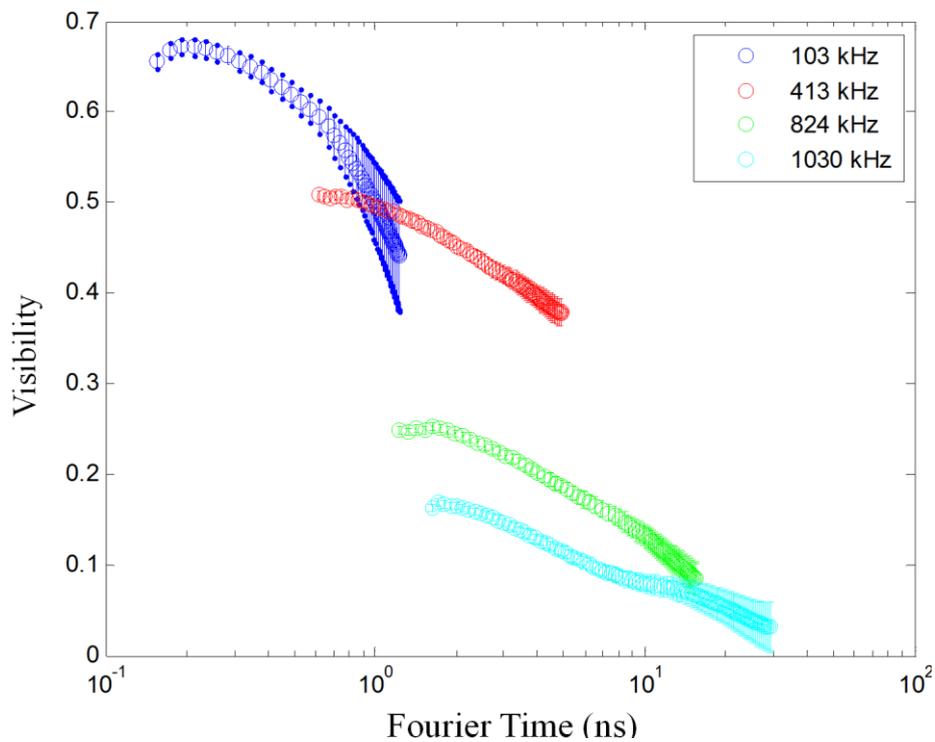

**Figure 5.** *Visibility as a function of Fourier time plotted for various MIEZE frequencies. The visibility shown here is the average visibility over the detector, with the frequency given in the legend.*

Figure 5 shows that the setup has a Fourier time range from ~100 ps to 30 ns. This resolution is comparable with other MIEZE setups previously reported. However, it is unique as it combines high energy resolution and time of flight, while not employing any material in the beam, thus providing good q resolution for SANS applications.

The MIEZE frequency shift (figure 6) due to the detector displacement was measured using a CASCADE detector and can be described well (within error) using eq. 5.

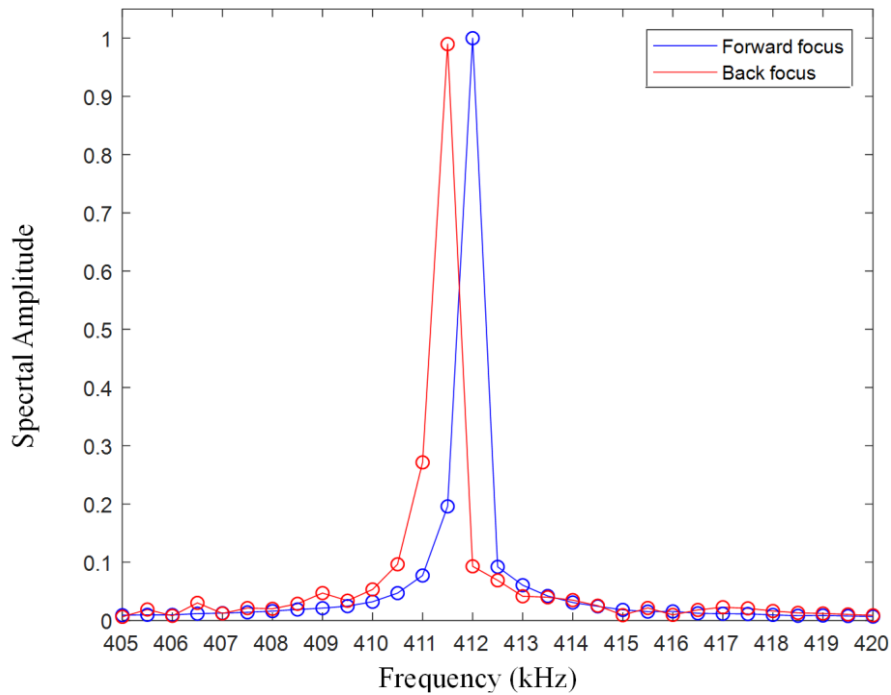

**Figure 6.** *MIEZE frequency shift due to a displacement of the detector (~ 1 cm). Since the distance between the last flipper and the detector is* 5.82 *m, the expected frequency shift according to eq. 5 is* 710 *Hz. The observed frequency shift is* 650 *Hz, with an uncertainty of 100 Hz.*

## 5. Discussion

The results demonstrate the feasibility of time of flight MIEZE on the Larmor instrument, with an energy resolution that is comparable to that of other MIEZE instruments. MIEZE on Larmor is unique as it offers a high energy resolution, time of flight and employs minimal material in the beam. The time of flight option yields a broad range of Fourier times, which are scanned, during each TOF pulse. However, this means that the scattering vector, q, is also scanned during each pulse. Thus for a single instrument setting the measured intermediate scattering function is on a certain contour of q-τ. This is not optimal for weak signals that occur at specific points in q space. The detector area and its efficiency are too small to measure dynamics of isotropic scatterers, thus for the time being anisotropic samples (i.e. coherent scattering) would be better candidates for TOF MIEZE on Larmor. Depolarising samples are also good candidates for MIEZE, as it is more challenging to characterize them using classical neutron spin echo. A resolution of 5 ns is sufficient for most magnetic samples [18], hence lower MIEZE frequencies (i.e. 100 kHz-400 kHz) are most applicable to these situations. Furthermore the 103 kHz (and lower) mode could also be used with the SANS $^3$He based detector on Larmor, which would enable MISANS measurements [35]. This $^3$He based detector would also yield a larger efficiency enabling measurements of quasi-elastic scattering from samples such as water. Due to the lower time resolution of this detector the modulation frequency would have to be below 100 kHz. The q range covered by this detector would be $0.003 - 0.7$ $\text{Å}^{-1}$.

Lower Fourier times could be achieved by reducing the modulation frequency further, by switching to π/2 MIEZE/MICE [36,37] or by placing the sample position closer to the detector. A combination of these methods would increase the Fourier time range of the instrument by another order of magnitude to the 10 ps domain. For high Fourier times the range cannot be increased significantly due to poor visibility. Figures 4 and 5 demonstrate that the visibility drops as the Fourier time is increased, the most plausible explanation for this is that the detector position did not perfectly coincide with the MIEZE focal point, as a result the two partial wavefunctions of the neutron would be slightly separated at the detector position. Due to the finite coherence length of the neutron, this separation results in decreased overlap and therefore interference between the two partial waves, leading to a reduction in visibility (see equation 4). Separation between the partial wavefunctions increases with Fourier time. This premise is supported by table 3, which shows that the measured frequency differs from the expected frequency, indicating that the detector position did not coincide with the focal point. The coherence length, which determines the sensitivity of the MIEZE signal to defocusing, is dictated by the wavelength resolution $\Delta\lambda/\lambda$, or the TOF distribution (the distribution of neutron flight times). At Larmor, this resolution is roughly 1% and is primarily governed by the size of the moderator and the distance between the moderator and the instrument.

Further optimisation of the experimental configuration is still possible through improvements of the analyser (reduce background) and detector system (efficiency) as well as the installation of background suppression elements such as evacuated flight paths and additional shielding.

## 6. Conclusion

Time of flight MIEZE for the Larmor instrument has successfully been tested. Modulation frequencies up to 1MHz have been achieved and a Fourier time range from ~0.1 ns to 30 ns was measured using a wavelength band from 4 Å to 11 Å. The visibility remains above 0.5 up to 1 ns Fourier time. Thus the Larmor instrument offers a unique time of flight MIEZE option, which combines high energy resolution, comparable to other MIEZE instruments.

**Acknowledgements**   We thank C. Pappas for the discussion and input regarding this publication. Experiments at the ISIS Neutron and Muon Source were supported by a beam time allocation RB1900030 [38] from the Science and Technology Facilities Council. This research was funded by a NWO groot grant No. LARMOR 721.012.102.